# Chaos and Formation of Structures in an Electron Flow with a Virtual Cathode in the Bounded Drift Tube

## Alexander Hramov

**Abstract**—The electron flow with a virtual cathode (VC) in the drift tube is investigated with the help of a 1.5-dimensional relativistic electromagnetic code. The existence of complex modes, including chaotic modes, is demonstrated. The dynamic nature of chaos in the considered system is revealed. Physical processes in the flow are investigated, and it is found that the initiation of chaotic dynamics of the electron flow with VC is related to the nonlinear interaction of structures formed in the system.

## INTRODUCTION

Oscillators operating on supercritical currents, or vircators [1], are promising sources of super-power microwave radiation. The generation mechanism in the vircator system is connected with formation in the electron flow of a region with the space-charge potential almost equal to the cathode potential—the so-called virtual cathode (VC); as a result of reflection of a part of electrons from the VC, the dual-flow state in the cathode–VC space is formed. Experimental and numerical data testify that the electron flow with VC has complex irregular dynamics. As early as in 1985, Brandt [2], considering the so-called turbotron, suggested that the dynamics of a device with VC is nonlinear. The studies of many groups of researchers revealed various manifestations of nonlinear dynamics of the electron flow with VC, for example, locking of the VC oscillations by external signals [3, 4] or chaotic behavior of a flow with VC [5]. In work [6], an assumption concerning the deterministic nature of chaos in a device with VC has been made.

At the same time, complex spatial–temporal dynamics in distributed systems is related to the interaction of emerging coherent spatial–temporal structures (see, for example, [7–9]). The knowledge of internal dynamics of the flow allows one to perform an efficient control of the VC oscillations and increase the efficiency and generated power in devices with VC, which are sufficiently important problems [6, 10].

In paper [11] devoted to investigation of a neutralized electron flow with VC in the Pierce diode, Anfinogentov has shown that the chaotic dynamics in this system is determined by the formation and nonlinear interaction of two electron bunches in the flow at each period of oscillations. These bunches are the VC itself and bunches formed as a result of the VC disintegration. These objects are the autostructures that emerge as a result of saturation of the Pierce and kinematic instabilities.

However, standard vircators usually operate without neutralization of the electron flow. Therefore, it is of special interest to clear up processes that cause chaotic dynamics in the electron flow without neutralization by the ion background with a supercritical current.

In this work, we present numerical results obtained with the help of a mathematical model of the vacuum microwave generator with VC, which is based on the self-consistent system of the Vlasov–Maxwell equations. Using the methods of nonlinear dynamics, we investigate the complex behavior of an electron flow with VC as well as the structure formation processes and their relation to the chaotic dynamics in the considered system.

## MATHEMATICAL MODEL

We investigate a system that represents a short-circuit section of a circular waveguide of length $L$ and radius $R$ placed in a strong magnetic field. The monospeed electron flow with the relativistic factor $\gamma = 2.3$ is injected inside the system through the cross section $z = 0$ (the plane of injection). We consider the one-dimensional motion of the flow along the $z$ axis under the assumption that the beam is focused by a strong longitudinal magnetic field.

Let us briefly consider the equations describing our system. By virtue of the axial symmetry and magnetization of the beam, the system of Maxwell equations degenerates into a system for three components $E_z$, $E_r$,

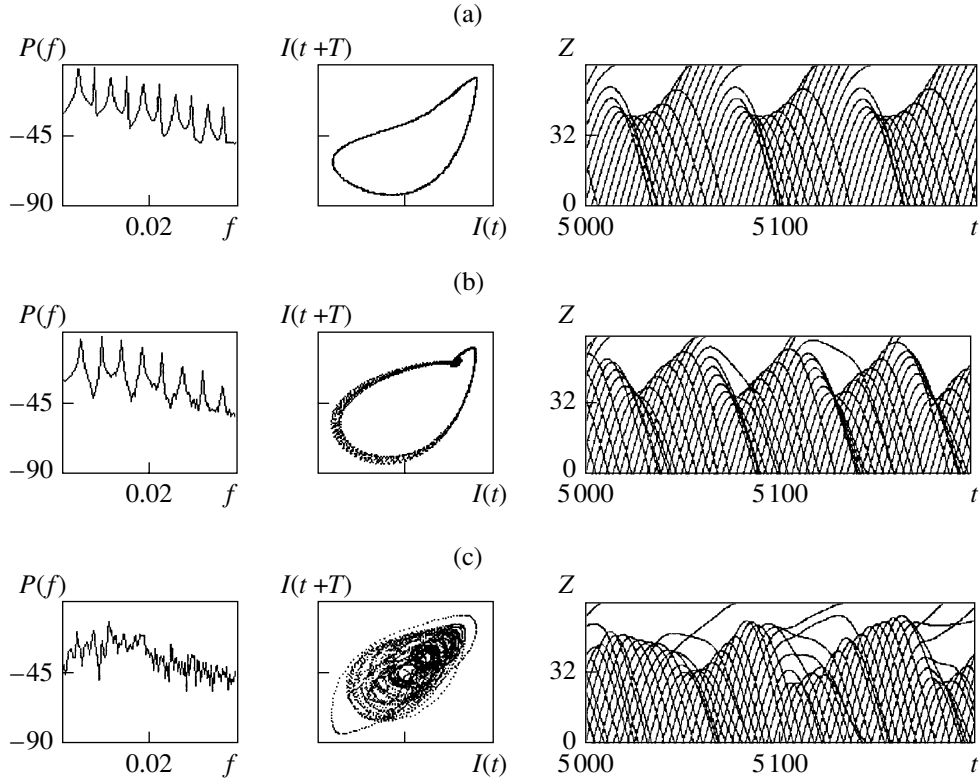

**Fig. 1.** Characteristics of the system oscillation for different operation modes: $\alpha$ = (a) 1.4, (b) 2.0, and (c) 4.0.

and $H_\theta$, which takes the following form in the cylindrical coordinate system:

$$\frac{1}{c}\frac{\partial E_z}{\partial t} = \frac{1}{r}\frac{\partial}{\partial r}(rH_\theta) - \frac{4\pi}{c}j_z, \quad (1)$$

$$\frac{1}{c}\frac{\partial E_r}{\partial t} = -\frac{\partial H_\theta}{\partial z}, \quad (2)$$

$$\frac{1}{c}\frac{\partial H_\theta}{\partial t} = \frac{\partial E_z}{\partial r} - \frac{\partial E_r}{\partial z}. \quad (3)$$

Solution to equations (1)–(3) satisfies the following boundary and initial conditions:

$$E_z|_{r=R} = E_r|_{\substack{z=0 \\ z=L \\ r=0}} = H_\theta|_{r=0} = 0, \quad (4)$$

$$E_z|_{t=0} = E_r|_{t=0} = H_\theta|_{t=0} = 0. \quad (5)$$

Dynamics of charged particles is described by the collisionless kinetic Vlasov equation

$$\frac{\partial f}{\partial t} + v_z \frac{\partial f}{\partial t} + eE_z \frac{\partial f}{\partial p_z} = 0, \quad (6)$$

where $f(t, z, p_z)$ is the distribution function of the beam electrons, $p_z$ is the relativistic momentum, and $v_z = p_z/m(1 + p_z^2/m^2c^2)^{1/2}$.

Solution of the Vlasov equation is based on the macroparticle method, which reduces (6) to a system of $N$ ($N$ is the number of macroparticles) ordinary differential equations

$$\frac{dv_{zk}}{dt} = (1 - v_{zk}^2)^{3/2} E_{zk}, \quad (7)$$

$$E_{zk} = \frac{1}{\Delta V_k} \iint E_z(r, z) h(|r - r_k|, |z - z_k|) r dr dz, \quad (8)$$

$$k = \overline{1, N}.$$

Here, $h$ is the normalized contribution function.

A conservative difference scheme for solving problem (1)–(8) can be constructed in accordance with the method set forth in papers [12–14].

## NONLINEAR DYNAMICS OF THE FLOW

For the specified geometry,[1] the main control parameter that affects the behavior of the considered system is the ratio $\alpha$ of the beam current to the ultimate vacuum current.

---

[1] The influence of the drift chamber geometry on the VC dynamics has been studied in detail in [15]; here, we will not consider this problem.

Numerical simulation has shown that, for $\alpha > 1$, a VC, which reflects a part of electrons, is formed in the flow, and the complexity of the VC oscillations increases together with the supercritical factor. Figure 1 presents the power spectra, projections of attractors of the time realizations of the beam current from the VC region restored according to the Takens method [16], and space–time diagrams of the electron flow in the drift space determined for different values of supercritical factor $\alpha$. Each line in the space–time diagrams is a trajectory of one charged particle.

For small $\alpha$ ($\alpha < 1.7$), regular oscillations of the relaxation type are established in the system (Fig. 1a). The power spectrum contains narrow peaks representing multiple harmonics of the fundamental frequency $\omega_0 \approx 2.6\omega_p$, where $\omega_p$ is the plasma frequency of the electron flow. The attractor projection corresponds to a one-time limiting cycle. When $\alpha$ increases, periodic oscillations decay, and two types of chaotic behavior arise sequentially as the supercritical factor increases. In the first case ($1.7 < \alpha < 3$), as one can see from Fig. 1b, the chaotic attractor arises on the basis of one unstable limiting cycle corresponding to the attracting set of periodic motions. In the second case ($\alpha > 3$, Fig. 1c), the phase portrait of oscillations is more uniform, and the attractor has a complex structure, which consists of a great number of unstable periodic orbits [17]; the power spectrum contains a large portion of noise and has no pronounced peaks. The form of the spectrum and the phase portrait indicates that an increased number of degrees of freedom is involved into the oscillatory motion in the system.

In order to analyze the attractor dimensionality, I calculated the correlation dimensionality using the Grassberger–Procaccia algorithm [18, 19]. The results confirm the deterministic nature of complex oscillations of the VC, which is caused by saturation of the attractor dimensionality that occurs when dimensionality $m$ of the space of imbedding increases. Such a behavior of the dimensionality indicates that chaotic modes are deterministic, because the correlation dimensionality of noise oscillations does not become saturated when $m$ increases [20].

When the supercritical factor is small, the dimensionality becomes saturated at small values of the dimensionality of the space of imbedding ($m = 3–4$). For the second chaotic mode, the space of imbedding has a substantially larger dimensionality, $m = 7–9$, which corresponds to the initiation of more complex oscillations of the VC. The number $n$ of the excited degrees of freedom in the system can be estimated by the upper bound of dimensionality $m_s$ of the system phase space, $n = m_s/2$ [21], where $m_s$ is equal to the dimensionality of the space of imbedding, at which the attractor dimensionality becomes saturated.

The analysis of the results presented above shows that only a small number of degrees of freedom is excited in the system, although the electron flow in the drift space represents a system with an infinite number of degrees of freedom. When $\alpha$ increases, the number of degrees of freedom involved in the oscillatory motion also increases. Nevertheless, this number is still rather small, which is indirect evidence that internal motions in the flow can be described with the help of a limited number of structures in the system.

## FORMATION OF STRUCTURES IN THE FLOW

The existence of coherent structures (see, for example, [7–9, 22] and the bibliographies in these works) indicates the presence of internal spatial–temporal dynamics in the distributed flow system. We extracted internal structures of the electron flow using the method of orthogonal decomposition [the Karhunen–Loéve (KL) expansion] [23, 24].

The problem of extraction of KL-modes is reduced to solving an integral equation

$$\int K(z, z^*)\Psi(z^*)dz^* = \lambda\Psi(z). \quad (9)$$

The kernel of the equation has the form

$$K(z, z^*) = \langle \xi(z, t)\xi(z^*, t)\rangle_t, \quad (10)$$

where $\langle\ldots\rangle_t$ denotes the time averaging. One can choose a set of space–time distributions of any physical quantity with zero mean value as functions $\xi(z, t)$. For the sake of convenience, I have chosen the values of the beam current $j_z(z, t)$ as functions $\xi(z, t)$. Eigenvalue $\lambda_n$ corresponding to the $n$th mode $\Psi_m$ is proportional to the energy contained in this mode. The quantity

$$W_n = \frac{\lambda_n}{\sum_i \lambda_i} \times 100\%$$

can be used as a measure of this energy.

Note that the KL expansion is optimal; in fact, eigenfunctions of problem (9) and (10) form a basis that fits this problem, since the rms error $\epsilon$ is minimal: $\epsilon = \min\langle\|\xi - \xi^N\|\rangle$, where $\xi$ is the exact solution, $\xi^N$ is the approximate solution, and $N$ is the basis dimensionality.

The table presents energies $W_n$ of the first ten modes for different values of the supercritical factor. In the regular mode ($\alpha = 1.4$), the first two modes contain about 90% of the flow energy ($W_1 + W_2 \sim 90\%$). When $\alpha$ increases, the spectrum of the mode energies broadens; the energy of the first mode is pumped over gradually to higher-order modes; and, for $\alpha \sim 4$, the main part of the energy is contained already in the first four modes ($W_1 + W_2 + W_3 + W_4 \sim 90\%$). The energies of the

Energy distribution over the KL modes for different VC oscillation modes

| Operation mode | Mode energy | | | | | | | | | |
|---|---|---|---|---|---|---|---|---|---|---|
| | $W_1$ | $W_2$ | $W_3$ | $W_4$ | $W_5$ | $W_6$ | $W_7$ | $W_8$ | $W_9$ | $W_{10}$ |
| Regular oscillations ($\alpha = 1.4$) | 64.08 | 23.67 | 7.33 | 2.73 | 1.54 | 0.44 | 0.15 | 0.05 | 0.01 | 0.00 |
| Weakly developed chaos ($\alpha = 2.0$) | 53.00 | 24.41 | 13.27 | 5.73 | 2.43 | 0.74 | 0.29 | 0.08 | 0.02 | 0.00 |
| Developed chaos ($\alpha = 4.0$) | 38.73 | 21.34 | 19.50 | 12.27 | 5.00 | 1.84 | 0.67 | 0.15 | 0.04 | 0.00 |

second, third, and fourth modes are almost equal. At the same time, in all operation modes (both periodical and with complex dynamics), the total number of modes with the energy above 1% of the total energy is small (about 5–7), which conforms with the results of Section 2 concerning a small number of degrees of freedom excited in the system.

Spatial patterns of modes have complex multihump form, and, as the mode number increases, they become more complicated and lose the symmetry that is characteristic of the higher-order mode.

Figure 2 displays the time realizations of amplitudes of the first two modes $A_1(t)$ (solid line) and $A_2(t)$ (dashed line) obtained as

$$A_n(t) = \int j_z(z,t)\Psi_n(z)dz. \quad (11)$$

When the supercritical factor is small (Fig. 2a), the modes behave rather regularly. One can select the time intervals, when a burst of amplitude $A$ is observed, which is coupled with formation and dynamics of VC in the flow, or when the mode amplitudes $A_1 \approx A_2 \approx 0$ (there are no structures in the flow; VC is open). Taking into account that oscillations of the first and second mode occur with a phase shift of $\pi/2$, we may assume that they describe the dynamics of the same spatial–temporal structure [25].

The developed chaos (Fig. 2b) is characterized by a complicated irregular mode dynamics; the structures permanently exist in the flow. It should be noted that, here, the mode amplitudes are lower than in the case of the regular operation mode. The cross-correlation function of processes $A_1$ and $A_2$ is strongly irregular with chaotic bursts, which allows us to relate each mode to its own structure in the flow.

Thus, a transition from regular to chaotic motions and sequential complication of chaotic operation modes is accompanied by an increase in the number of the flow structures, and the mode energies draw together.

Analysis of space–time diagrams (see Fig. 1) allows one to establish a relationship between the KL-modes obtained using orthogonal decomposition and physical processes that accompany the complication of the VC dynamics. As one can see from Fig. 1a, there exists only one structure (VC) in the regular operation mode (the time interval of its existence within each oscillation period—from the formation to disintegration—coincides with the burst of the mode amplitudes in

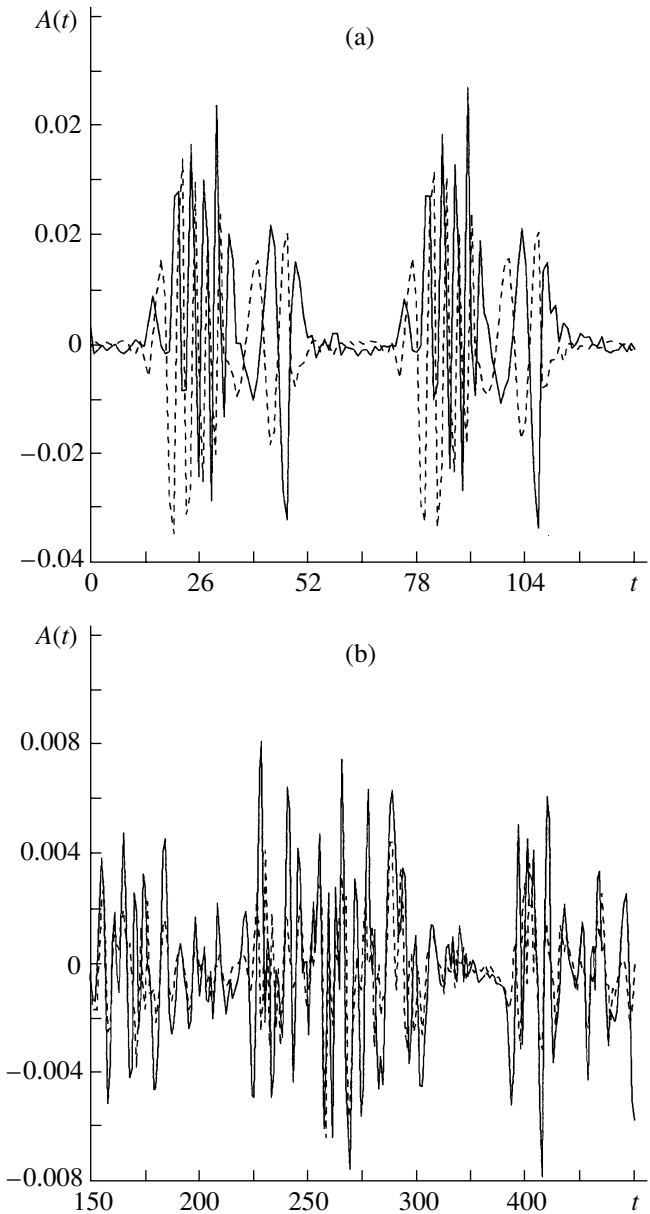

**Fig. 2.** Amplitudes of the (solid line) first and (dashed line) second harmonics of the Karhunen–Loéve modes vs. time for different operation modes: $\alpha =$ (a) 1.4 and (b) 4.0.

Fig. 2a). The main portion of the energy of oscillatory motion in the system falls on higher-order modes $\Psi_1$ and $\Psi_2$; their behavior describes the dynamics of this structure.

As $\alpha$ increases (due to kinematic instability of the electron flow in the varying field of the opening VC), the secondary structures are formed in the flow; the energy increases gradually together with $\alpha$ due to the energy transfer from the higher-order mode. As can be seen from Fig. 1c, a well-developed VC is absent in the flow in the case of developed chaos. This situation may be interpreted as the existence in the flow of several typical structures described by KL-modes. Each of these structures is considered as a proper VC; i.e., as the plane reflecting charged particles, which is located at a certain distance from the injection plane and corresponding to the charge density maximum in the drift space. Interaction of these structures (VC) through the reflection of a part of the flow from each of them strongly affects the formation of other structures and ensures, due to this fact, a distributed internal feedback in the flow. This process is equivalent to the formation of several feedback loops with different time delays, which may explain strongly irregular dynamics of the system in the case of large supercritical factor $\alpha$.

## CONCLUSION

The electron flow with VC and without neutralization by the ion background in the bounded drift tube demonstrates different types of nonlinear oscillations. For a small supercritical factor ($\alpha < 3$), a low-dimensional chaos is established in the system. When $\alpha$ increases, the system demonstrates the developed chaos. In the regular operation mode, there is only one structure (VC) in the flow; the VC can be described by the dynamics of the first two KL modes; more than 85% of the energy of oscillatory motion falls on these modes. The chaotic dynamics can be explained by the formation and interaction of several coherent structures in the electron flow with a smoother distribution of the energy over modes.

The understanding of physical processes in the considered system with VC allows one to develop efficient methods to control the system by affecting the structure formation in the electron flow with VC. Apparently, among these mechanisms that affect the formation of structures in the flow is the introduction into the system with VC of an external or internal feedback [26–28], which enables one to suppress the formation of secondary structures by means of a preliminary modulation of the electron flow.

## ACKNOWLEDGMENTS

I am grateful to V.G. Anfinogentov for attention to this work, numerous discussions, and helpful critical remarks.

This work was supported by the Russian Foundation for Basic Research, project no. 96-02-16753.